\documentclass[prl,twocolumn,showpacs,superscriptaddress, floatfix, reprint]{revtex4-1}

\usepackage[OT1]{fontenc}
\usepackage[usenames,dvipsnames]{color}
\usepackage[latin1]{inputenc}
\usepackage[english]{babel}
\usepackage{graphicx}
\usepackage{color}
\usepackage{amssymb,amsmath}
\usepackage[Gray,squaren]{SIunits}
\usepackage{xspace}
\usepackage{upgreek}
\usepackage{ulem}
\usepackage{epstopdf}
\usepackage{hyperref}
\newcommand{\ud}{\mathrm{d}}
\newcommand{\vbar}{\bar{v}}
\newcommand{\Iin}{I_0}
\newcommand{\Iout}{I_t}

\newcommand{\Isat}{I_{\mathrm{sat}}}
\newcommand{\Ein}{E_0}
\newcommand{\Eout}{E_t}
\newcommand{\Esca}{E_s}

\newcommand{\ie}{i.e.\@\xspace}
\newcommand{\eg}{e.g.\@\xspace}

\newcommand{\eq}[1]{Eq.~\eqref{eq:#1}}
\newcommand{\eqs}[2]{Eqs.~\eqref{eq:#1} and~\eqref{eq:#2}}

\newcommand{\fig}[1]{Fig.~\ref{fig:#1}}

\newcommand{\erfc}{\operatorname{erfc}}
\newcommand{\im}{\operatorname{Im}}
\newcommand{\re}{\operatorname{Re}}
\newcommand{\maxop}{\operatorname{max}}
\newcommand{\D}{\mathrm{d}}

\makeatletter
\newcommand*{\balancecolsandclearpage}{%
  \close@column@grid
  \clearpage
  \twocolumngrid
}

\begin{document}

\title{Cooperative Emission of a Pulse Train in an Optically Thick Scattering Medium}
\author{C. C. Kwong}
\affiliation{School of Physical and Mathematical Sciences, Nanyang Technological University, 637371 Singapore,
Singapore.}
\author{T. Yang}
\altaffiliation[Present address: ]{National Institute of Metrology, Beijing, China}
\affiliation{Centre for Quantum Technologies, National University of Singapore, 117543 Singapore, Singapore}
\author{D. Delande}
\affiliation{Laboratoire Kastler Brossel, UPMC, CNRS, ENS, Coll\`ege de France, 4 Place Jussieu, F-75005 Paris, France}
\author{R. Pierrat}
\affiliation{ESPCI ParisTech, PSL Research University, CNRS, Institut Langevin, 1 rue Jussieu, F-75005 Paris, France}
\author{D. Wilkowski}
\email{david.wilkowski@ntu.edu.sg}
\affiliation{School of Physical and Mathematical Sciences, Nanyang Technological University, 637371 Singapore,  Singapore.}
\affiliation{Centre for Quantum Technologies, National University of Singapore, 117543 Singapore, Singapore}
\affiliation{MajuLab, CNRS-University of Nice-NUS-NTU International Joint Research Unit UMI 3654, 117543 Singapore, Singapore}

\begin{abstract}
An optically thick cold atomic cloud emits a coherent flash of light in the forward direction when the phase of an
incident probe field is abruptly changed. Because of cooperativity, the duration of this phenomena can be much shorter
than the excited lifetime of a single atom. Repeating periodically the abrupt phase jump,
we generate a  train of pulses with short repetition time, high intensity contrast and high efficiency.
In this regime, the emission is fully governed by cooperativity even if the cloud is dilute.
\end{abstract}



\maketitle

The seminal work on superradiance of R. Dicke in 1954 has opened up tremendous interest in studying cooperative emission
of electromagnetic radiation from an ensemble of radiative dipoles (see~\cite{dicke1954coherence} for the original
proposal,
\cite{gross1982superradiance,brandes2005coherent} for reviews and
\cite{paradis2008observation,das2008quantum,scully2009super,ariunbold2010observation,bohnet2012steady,ott2013cooperative,mlynek2014observation}
for recent related works). In his original proposal, R. Dicke considered an ensemble of $N$ excited two-level atoms
confined inside a volume smaller than $\lambda^3$, where $\lambda$ is the transition wavelength. In this context, a
macroscopic polarization is built up in the medium upon incoherent spontaneous emission. This Dicke superradiance
mechanism leads to the coherent emission of an intense pulse with a decay time, $\tau_{D}=(N\Gamma)^{-1}$, that is
shortened by a factor of $N^{-1}$ with respect to the atomic excited state lifetime, $\Gamma^{-1}$. For practical
implementation in the optical domain, the Dicke model was extended to media with volume larger than $\lambda^3$
\cite{rehler1971superradiance,bonifacio1975cooperative,gross1982superradiance}. In those cases, the
propagation of the electromagnetic field in the medium and the spatial mode density must be taken into account.
If the medium is dense, \ie,
$\rho\lambda^3\gg1$, where $\rho$ is the radiator spatial density, it still exhibits the main feature of the Dicke
superradiance, namely, the emission of a short pulse after some
delay~\cite{skribanowitz1973observation,marek1979observation,paradis2008observation,ariunbold2010observation}. It was,
however, pointed out in \cite{rehler1971superradiance}, that the superradiant pulse
decay time should be corrected as $\tau=\tau_{D}/\mu$. $\mu<1$ is a geometrical factor corresponding to the solid
angle subtended by the superradiant emission~\cite{gross1982superradiance,bonifacio1975cooperative}.

For a dilute scattering medium, \ie, $\rho\lambda^3\ll1$, the Dicke superradiance
mechanism does not occur~\cite{Note1}. Nevertheless, an optically thick medium
 driven by a coherent incident field shares interesting similarities with
Dicke superradiance; here, the cooperativity factor $N\mu$ is replaced by the optical
thickness of the medium~\cite{friedberg1971superradiant,friedberg1976superradiant}. Once a driving coherent
field is abruptly switched off, like in a free induction decay (FID) experiment
\cite{hahn1950nuclear,brewer1972optical,foster1974interference,toyoda1997optical,Shim2002,Wei2009,chalony_coherent_2011,
kwong2014},
a short coherent
cooperative flash of light is emitted in the forward direction. The flash duration is inversely proportional
to the optical thickness and the bare linewidth of the transition~\cite{chalony_coherent_2011}.
A similar phenomenon occurs for the optical precursor, \ie, when the driving coherent
field is abruptly switched on \cite{jeong2006}.

\begin{figure}
   \begin{center}
      \includegraphics[width =\linewidth]{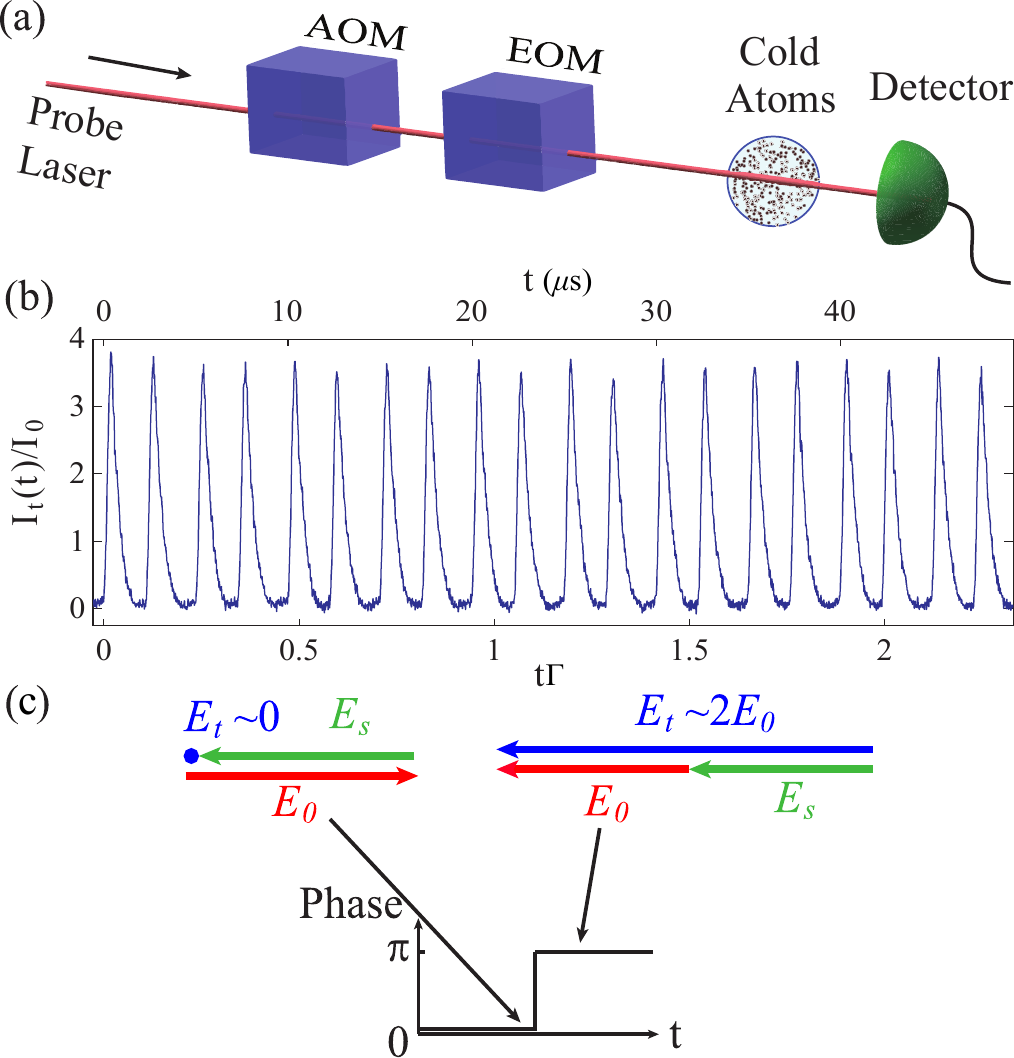}
      \caption{(color online).  (a) Experimental setup: a laser beam is sent through an acousto-optic modulator (AOM),
      which may be used to switch
      on and off the incident beam, followed by an electro-optic modulator (EOM), which abruptly changes the phase of
      the incident probe field, $\Ein$.  (b) A pulse train
      generated at a repetition time of $T_R=0.12\Gamma^{-1}$ by a periodic abrupt phase change of $\pi$.
      Here, the probe laser is at resonance.
(c) Electric fields just before and just after an abrupt phase jump of $\pi$ are
represented
      schematically in the complex plane. Before the phase jump,
      the forward scattered field $\Esca$ destructively
      interferes with $\Ein$. After the phase jump, they constructively  interfere.
      The transmitted field is denoted by $\Eout$.}
      \label{fig:setup}
   \end{center}
\end{figure}

In a coherently driven medium, the incident probe frequency can be detuned with respect to the atomic resonance,
leading to a nontrivial phase rotation
of the cooperatively emitted field (see~\cite{kwong2014} in the optical domain and
\cite{helisto1991,shakhmuratov2015,antonov2015} for $\gamma$-ray pulses
in M{\"o}ssbauer spectroscopy experiments).
In this Letter,
we report the generation of high repetition rate and high intensity contrast pulse trains in an optically
thick cold dilute atomic ensemble using
the setup schematically shown in~\fig{setup}(a). An example of a pulse train, generated in our experiment
by periodically changing the
probe phase, is shown in~\fig{setup}(b). As a consequence of cooperative emission, the repetition time
$T_R$ of the pulse train can be shorter than the atomic excited state lifetime, $\Gamma^{-1}$.
Moreover, we show that at high repetition rate, the single atom fluorescence is quenched.
This constitutes a rather counterintuitive result where the emission in free space is fully governed
by cooperativity, in contrast with the usual situations where it is enhanced by a cavity
surrounding the medium~\cite{carmichael1991}.

The scattering medium is a cloud of laser-cooled \textsuperscript{88}Sr atoms (see
\cite{tao2015} and \cite{kwong2014} for the details of the cold atoms production line). The
ellipsoidal
shape of the cold cloud has an axial radius of $\unit{240(10)}{\micro\meter}$ and an equatorial radius of
$\unit{380(30)}{\micro\meter}$, with peak density around $\unit{4.6\times 10^{11}}{\rpcubic{\centi\meter}}$
for a total of $2.5(5)\times 10^8$  atoms. $\lambda=\unit{689}{\nano\meter}$ is the
wavelength associated to the $^1\!S_0\rightarrow\,^3\!P_1$ intercombination line (bare linewidth of
$\Gamma/2\pi=\unit{7.5}{\kilo\hertz}$) used in this experiment. $\rho\lambda^3=0.15$, which puts us in the dilute
regime. The temperature of the cold gas is
$T=\unit{3.3(2)}{\micro\kelvin}$. We get $k\vbar=3.4\Gamma$, indicating a significant
Doppler broadening of the narrow intercombination line. $k=2\pi/\lambda$ is the wave vector of the transition, and
$\vbar$ is the rms velocity of the gas.
The optical thickness depends strongly on the
temperature. We measure $19(2)$ along the equatorial axis at resonance.

A 150~$\mu$m diameter probe laser beam, tuned around the intercombination line, is sent through the cold
atomic gas along an equatorial axis. The probe power is $\unit{400(40)}{\pico\watt}$, corresponding to $0.45(5)\Isat$
(${\Isat=3~\mu\text{W/cm}^2}$).  We measure the forward transmitted intensity of the probe
using a photodetector, integrating over the transverse dimensions of the transmitted beam.  We apply a bias
\unit{1.4}{G}  magnetic field along the beam polarization during the probing phase, making the
atom an effective two-level system on the  $^1\!S_0, m=0\rightarrow\,^3\!P_1, m=0$ transition.

The ellipsoidal shape of the cloud is modeled by a slab geometry, so that the coherent transmitted electric field, in
the frequency domain, is given by
\begin{equation}
   \Eout(\omega)=\Ein(\omega)\exp\left[i\frac{n(\omega)\omega L}{c}\right].
   \label{eq:field_trans}
\end{equation}
In the above equation, $n(\omega)$, $\Ein$, $c$, and $L$ are the complex effective refractive index, the incident
optical field, the speed of light in vacuum, and the slab thickness along the laser beam, respectively. For a
dilute medium, $n(\omega) = 1 + \rho\alpha(\omega)/2$~\cite{Hetch1974}, with the two-level atomic polarizability,
\begin{equation}
   \alpha(\omega)=-\frac{3\pi\Gamma c^3}{\omega^3}\frac{1}{\sqrt{2\pi}\vbar}\int^{+\infty}_{-\infty} dv
      \frac{\textrm{exp}\left(-v^2/2\vbar^2\right)}{\delta-kv+i\Gamma/2}.
   \label{eq:polarizability}
\end{equation}
 $\delta=\omega-\omega_0$ is the detuning of the probe laser frequency $\omega$ with respect to the bare atomic
resonance
frequency, $\omega_0$. The effect of Doppler broadening is included in the polarizability by averaging over the thermal
Gaussian distribution of the atomic velocity $v$ along the beam propagation direction. The transmitted intensity
$\Iout(t)$ is computed following~\cite{kwong2014}, and by performing an inverse Fourier transform. We define, for given $\delta$ and $\vbar$,
the optical thickness $b_{\vbar}(\delta)$ and the relative phase $\theta_{\vbar}(\delta)$
between the transmitted and the incident fields by
\begin{align}
   b_{\vbar}(\delta) = \,&\frac{2\omega}{c}\im[n(\omega)]L,\nonumber\\
   \theta_{\vbar}(\delta) =\,& \frac{\omega}{c}\re[n(\omega)-1]L.
   \label{eq:b_and_theta}
\end{align}

The transmitted field $\Eout$ results from the interference between the incident field $\Ein$
and the field scattered in the forward direction $\Esca$,
\begin{equation}
   \Eout = \Ein + \Esca.
   \label{eq:field_relation}
\end{equation}
For effective two-level atoms, we can drop the vectorial nature of the electric fields and represent them as
scalar quantities.  Because of the noninstantaneous response time of the medium, the coherent scattered field in the
forward direction is a continuous function across the abrupt change of the incident field.  In a FID experiment
where the incident field is abruptly switched off at $t=0$, the intensity of the transmitted field at $t=0^+$ is a
direct measurement of the forward scattered intensity in the stationary regime. Its properties are studied in detail in
~\cite{chalony_coherent_2011,kwong2014}. In particular, the intensity of the forward scattering is bounded
by 4 times the incident intensity
(``superflash effect'')~\cite{kwong2014}.
The temporal evolution of the transmitted field, after the abrupt switch off of the incident field, is
not a simple function having only one characteristic decay rate~\cite{chalony_coherent_2011}. However, we get a clear
physical
insight by considering only the initial decay time (at $t=0^+$), which takes a simple analytical expression (see
Supplemental Material
~\cite{Note2}):
\begin{multline}
   \tau_{\vbar}(\delta) =
   \left|\frac{\Iout(t=0^+)-\Iout(t=\infty)}{d\Iout/d t(t=0^+)}\right| \\
   = \frac{2}{\Gamma b_0(0)}\frac{1+\exp(-b)-2\exp(-b/2)\cos(\theta)}{1-\exp(-b/2)\cos(\theta)}
   \label{eq:decay_time}
\end{multline}
where $b\equiv b_{\vbar}(\delta)$ and $\theta\equiv
\theta_{\vbar}(\delta)$. In \eq{decay_time},
$b_0(0)$ is the optical thickness at resonance and zero velocity. It is linked to $b_{\vbar}(0)$ by
$b_{\vbar}(0)=b_0(0)g(k\vbar/\Gamma)$ where $g(x) = \sqrt{\pi/8}
\exp\left(1/8x^2\right)\erfc\left(1/\sqrt{8}x\right)/x$~\cite{chalony_coherent_2011}.

\begin{figure}
   \begin{center}
      \includegraphics[width = \linewidth]{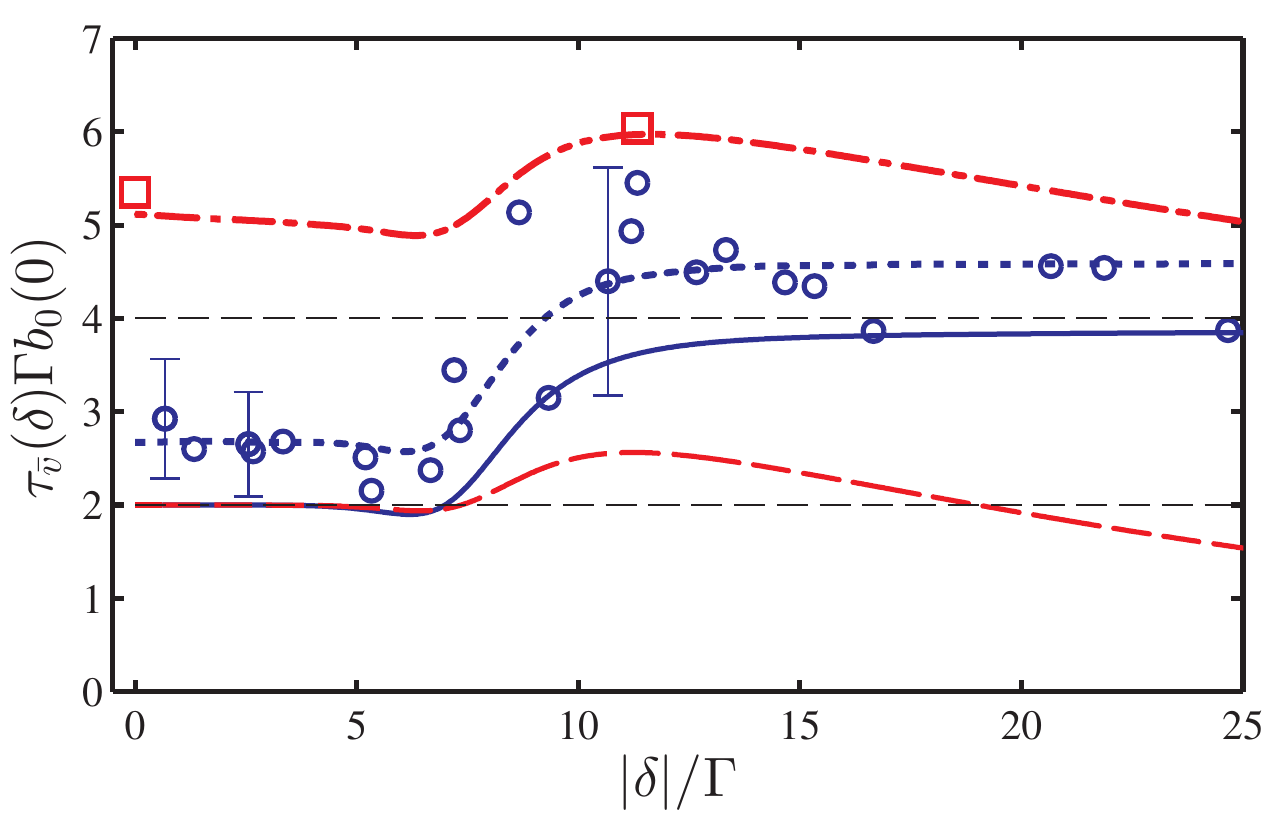}
      \caption{(color online). Initial decay time of the coherent flash versus the probe detuning at $k\vbar/\Gamma =
      3.4$. The zero temperature resonant optical thickness is $b_0(0)=120$. The blue open circles and plain curve are,
      respectively, the experimental data points and the theoretical curve for an abrupt switch off of the probe.  The
two horizontal black dashed lines give the theoretical
      predictions at resonance and at large detuning. The red squares and dashed curve are the
      experimental data points and theoretical prediction for an abrupt phase jump
       of $\pi$. The blue dotted
      line and the red dash-dotted line are numerical predictions taking into
      account the finite response time of the experimental scheme (see text for more
      details).}
      \label{fig:tau}
   \end{center}
\end{figure}

For small optical thickness, \eq{decay_time} reduces to $\tau_{\vbar}(0) = g(k\vbar/\Gamma)/\Gamma$ at resonance. It is
shorter than $\tau_0(0) = 1/\Gamma$ due to the dephasing effect from the motion of the atoms. This has already been observed
experimentally [see Fig.~3(b) of~\cite{chalony_coherent_2011} where the transition is Doppler broadened, and
Fig.~5 of~\cite{Shim2002} where Doppler broadening can be ignored]. In our
experiments, $g(k\vbar/\Gamma)\approx0.16$; thus, for $b_{\vbar}(0)=19(2)$, we get $b_0(0)=120(10)$.
A direct measurement gives a slightly smaller value, $b_0(0)=95(5)$
(see Supplemental Material~\cite{Note2}).
The expression of
$\tau_{\vbar}$, given by \eq{decay_time}, simplifies to $\tau_{\vbar}(0) =
2[b_0(0)\Gamma]^{-1}$ at resonance ($b\gg 1$, $\theta=0$) and to $\tau_{\vbar}(\pm\infty) =
4[b_0(0)\Gamma]^{-1}$ far from resonance ($b=0$ and $\theta=0$). The solid blue curve in~\fig{tau} is a plot of $\tau_{\vbar}(\delta)$ for
$k\vbar/\Gamma = 3.4$. $\tau_{\vbar}$ has a weak dependence on $\delta$ and
$\vbar$; it depends mainly on $b_0(0)$, which can be much larger than the optical
thickness $b_{\vbar}(0)$ \emph{seen} by a resonant probe at nonzero
temperature. This strongly reduces the lifetime of the forward scattered field with respect to
the atomic lifetime, $\Gamma^{-1}$.
Equation~\eqref{eq:decay_time} has a rather
simple physical interpretation: the second term represents the geometrical properties of the propagation inside the
medium (change in amplitude and phase shift) while the term $2/\Gamma b_0(0)$ represents the collective behavior of
all excited dipoles. It does not depend on the atomic velocity, but only on the atomic density integrated along the
laser direction, because there is no Doppler effect for photons scattered in the forward direction. Similarly, it does
not depend on the detuning because all dipoles  decay with the same rate
$\Gamma$ independently of the detuning.

The FID experiment is performed using an AOM as a light
switching device [see
Fig~\ref{fig:setup}(a)]. The experimental data points, represented by blue open circles in~\fig{tau}, are in
reasonable
agreement with the theoretical prediction. The
evaluation of $d \Iout/d t(t=0^+)$ is performed on a short temporal window ($\sim
\unit{200}{\nano\second}$) after switching off the incident probe.
While the flash signal has a good signal to noise ratio [see~\fig{setup}(b)],
the resulting $d \Iout/d t(t=0^+)$ values from this analysis are noisier. This leads to the large statistical errors for $\tau_{\vbar}$.
The slight positive systematic error, also associated to the determination of $d \Iout/d
t(t=0^+)$, comes from the finite response time of our experimental scheme,
of the order of $\unit{40}{ns}\approx (500\Gamma)^{-1}$. To check the latter statement, we
use~\eqs{field_trans}{polarizability} to numerically compute $\Iout(t).$
$\Ein(\omega)$ in~\eq{field_trans} is determined from the measured time evolution of the incident intensity.
We then apply, on the numerical signal, the same procedure used experimentally to extract $\tau_{\vbar}$,
resulting in an excellent agreement with the experimental data (see \fig{tau}).

Instead of a FID experiment, we now consider an
abrupt jump of the phase of the incident field by $\pi$ [see~\fig{setup}(c)], at constant incident intensity.
The initial decay time $\tau_{\vbar}$ becomes (see Supplemental Material~\cite{Note2}):
\begin{equation}
   \tau_{\vbar}(\delta) = \frac{4}{\Gamma b_0(0)} \frac{1-\exp(-b/2)\cos\theta}{2-\exp(-b/2)\cos\theta}.\label{eq:phase_
decay_time}
\end{equation}
We plot this expression as the red dashed line in~\fig{tau}. If the $\pi$ phase jump occurs at $t=0$, according to \eq{field_relation},
we have $\Eout(t=0^+)=-\Ein(t=0^-)+\Esca(t=0^-)$. To observe the largest possible amplitude of the
transient field, we choose the probe frequency
detuning such that
the interference between $\Ein(t=0^-)$ and $\Esca(t=0^-)$ is destructive. This condition is necessarily fulfilled when
the incident field is at resonance. If $b_{\vbar}(0)\gg 1$, $|\Esca(t=0^-)|\simeq|\Ein|$, so we
expect a coherent flash with a peak intensity, $\Iout=4\Iin$. The destructive interference condition may also happen at
a nonzero detuning if the phase rotation experienced by $\Esca$
is large enough, for example if $b_{\vbar}(0)\gg 1$. In our experiment,
this situation occurs at $|\delta|=11.3\Gamma$ (\ie, superflash
regime~\cite{kwong2014}). In this context, $|\Esca(t=0^-)|\simeq 1.8|\Ein|$; thus, the flash has a peak intensity
$\Iout\simeq(1+1.8)^2\Iin\simeq7.8\Iin$. This value is
slightly below the maximum value $9\Iin$ allowed by energy conservation,
achievable at larger optical thickness.

The phase jump is performed using an EOM placed on the probe laser path
[see~\fig{setup}(a)]. The EOM is driven by a high voltage controller and has a slew rate
$\sim 2.3\pi~\text{rad }\mu\text{s}^{-1}$. The two experimental values (red squares), corresponding to $\delta=0$ and
$|\delta|=11.3\Gamma$, are shown in \fig{tau}. They are systematically higher than the theoretical
prediction for an abrupt phase shift change because of the
response
time of the EOM driver. Similarly to the FID experiment, we use the experimentally
measured EOM driver output
to numerically compute the $\Iout(t)$ signal. The resulting values of the decay time (red
dash-dotted line in \fig{tau}) agree with the experimental ones.

We now analyze the cooperative
emission when a square periodic $\pi$ phase jump is applied.
We observe a pulse train
 with a repetition time $T_R$ [see an example
in~\fig{setup}(b)] limited by the relaxation time of the system. The
cooperative emission in the forward direction dramatically decreases the repetition
time below the atomic excited state lifetime.

\begin{figure}
   \begin{center}
      \includegraphics[width = \linewidth]{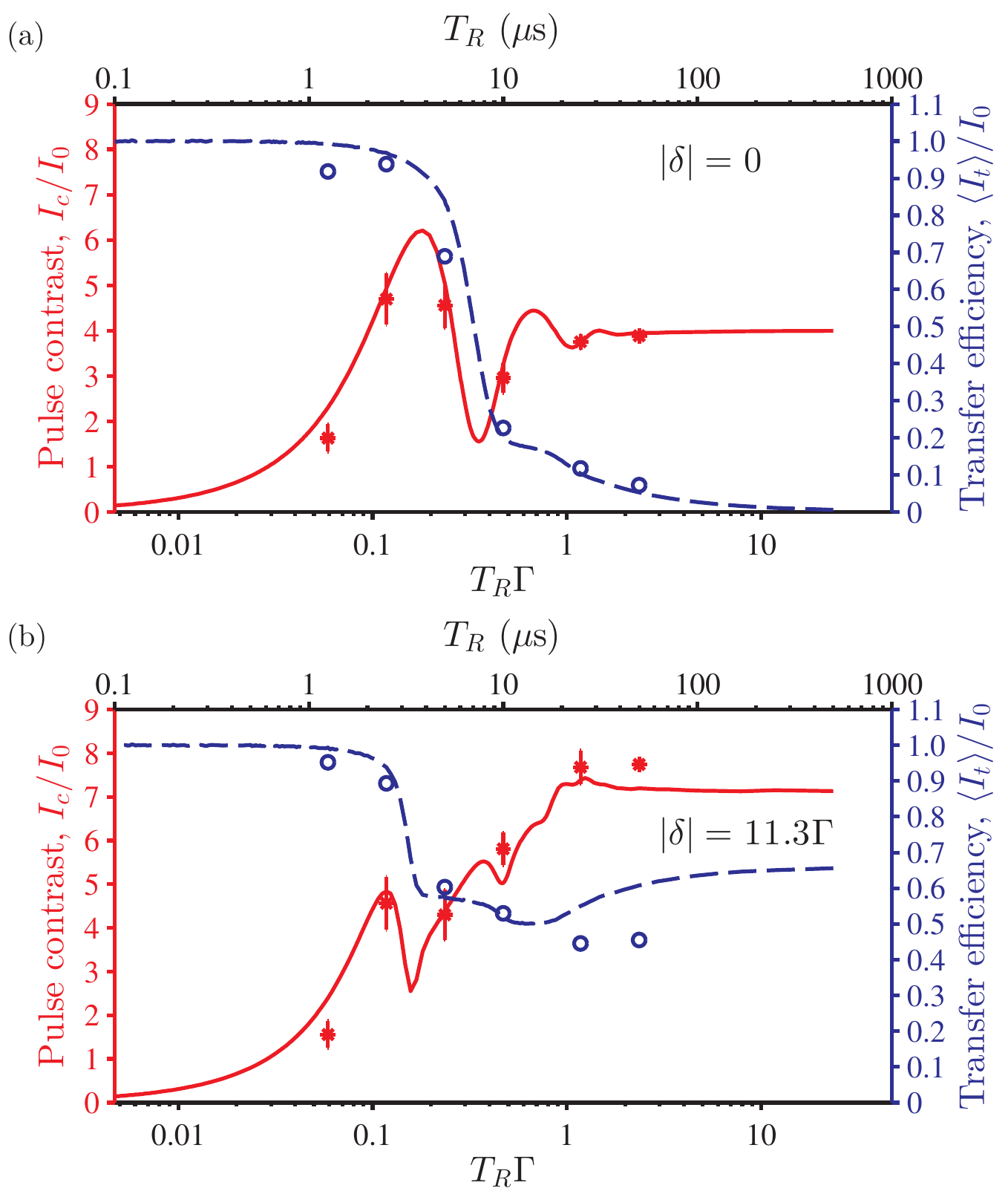}
      \caption{(color online). Figures of merit of the generated pulse train  (a) at resonance and (b) at
      $|\delta| = 11.3\Gamma$. The red solid and  blue dashed curves are the theoretical predictions for the  intensity contrast
      $I_c / \Iin$ and transfer efficiency $\langle \Iout \rangle/\Iin$, respectively. The red
      dots and blue open circles are the corresponding experimentally measured values.}
      \label{fig:figure_merit}
   \end{center}
\end{figure}

Bringing the probe on resonance, we plot in~\fig{figure_merit}(a) (red dots and solid curve)
the intensity contrast
$I_c$ of the pulse train. We define $I_{c}=\maxop\{\Iout\}-\langle \Iout \rangle$ as the
difference between the maximum intensity $\maxop\{\Iout\}$ and the mean intensity, $\langle \Iout
\rangle=1/T_R\int_{T_R}\Iout(t)d t$. We observe an excellent agreement between the experiment and the
theoretical prediction of~\eqs{field_trans}{polarizability}. At long
repetition time, \ie, $T_R\gg\Gamma^{-1}$, the system reaches its steady state
before every phase jump.  Hence, we measure $I_c \simeq 4\Iin-\langle \Iout \rangle\simeq 4\Iin$. We
note that $\langle \Iout \rangle\simeq 0$ [see the blue open circles and dashed curve
in~\fig{figure_merit}(a)] and  most of the incident power is scattered out by single atom fluorescence events. In the $\tau_
{\vbar}\lesssim T_R\lesssim\Gamma^{-1}$
intermediate regime, $I_c$ oscillates and can reach a larger value. Moreover, the mean intensity $\langle \Iout \rangle$
rapidly increases to its maximal value, $\Iin$. Here, the incident power is almost perfectly transferred to the pulse
train. This interesting result can be understood
considering cooperativity in forward scattering. Indeed, its characteristic relaxation time scales like
$[b_0(0)\Gamma]^{-1}$.
Therefore, for $b_0(0)\gg 1$, coherent processes relax much
faster than single atom fluorescence events. The latter are quenched, leading to the good figure of merit at
a repetition time shorter than $\Gamma^{-1}$. In other words, the emission from the atoms is governed by cooperativity.
For $T_R<\tau_{\vbar}$, the repetition rate is faster than any time scale of the
atomic ensemble. Even though the probe power is fully transmitted, the contrast $I_c$ tends to zero.

At detuning $|\delta| = 11.3\Gamma$, for long repetition time ($T_R\gg\Gamma^{-1}$), the pulses have a higher contrast,
$I_c\simeq(1+1.8)^2\Iin-\langle \Iout \rangle\simeq 7.1\Iin$ [see~\fig{figure_merit} (b)].
The large value
of the mean intensity, namely, $\langle \Iout \rangle\simeq 0.7\Iin$, is due to
the
small optical thickness, $b_{\vbar}(\delta)=0.4$. Hence, most of the transmitted
power is in
a continuous transmission mode and not in the pulse train. At the intermediate repetition time ($\tau_{\vbar}\lesssim
T_R\lesssim \Gamma^{-1}$), the pulse contrast and the figure of merit are not as good as in the resonant case.

To conclude, we generate pulse trains of short repetition time using cooperative forward emission in an optically thick
scattering medium. We can almost completely transfer the incident power into the high intensity contrast pulse train,
quenching the single atom fluorescence. This means that, in free space, the cooperativity effect can dominate
emission from a dilute
atomic gas. The decay time of the pulses
also weakly depends on the temperature of the gas and on the probe detuning. An interesting extension of this study
could be to look for quantum signatures in the cooperative emission.

Finally, we employ the narrow intercombination line of strontium as a proof of principle, where the time scales are of
the order of microseconds. For future practical applications, such as a high contrast pulse generator, shorter
repetition times in the picosecond or subpicosecond regime should be attainable. For this purpose, one has to use
scattering media with higher optical thickness and/or shorter transition lifetime. The fact that
cooperativity is robust over thermal dephasing means that we can also use a hot vapor of rubidium
[$b_{\vbar}(0) \approx 600$ at $\unit{110}{\celsius}$~\cite{weller2011absolute}]. We can also use
condensed matter systems, \eg, a samarium doped fiber [$b(0)\approx
100$~\cite{SmTrans}], which allows us to bring this technique into the 1.55~$\mu$m
telecommunication
band.

The authors thank M.~Pramod, F.~Leroux, and K.~Pandey for technical support and fruitful discussions. C.~C.~K. thanks
the CQT and ESPCI institutions for funding his trip to Paris.  This work was supported by CQT/MoE funding, Grant No.
R-710-002-016-271.  R.~P. acknowledges the support of LABEX WIFI (Laboratory of Excellence ANR-10-LABX-24) within
the French Program ``Investments for the Future'' under reference ANR-10-IDEX-0001-02 PSL$^{\ast}$.
%

\newcommand{\noopsort}[1]{}\providecommand{\noopsort}[1]{}\providecommand{\singleletter}[1]{#1}%


\clearpage

\twocolumngrid
\setcounter{equation}{0}
\setcounter{figure}{0}
\setcounter{table}{0}
\setcounter{page}{1}
\renewcommand{\thepage}{S\arabic{page}}
\renewcommand{\thesection}{S\arabic{section}}
\renewcommand{\thetable}{S\arabic{table}}
\renewcommand{\thefigure}{S\arabic{figure}}
\renewcommand{\theequation}{S\arabic{equation}}

\onecolumngrid

\phantomsection
\section*{Supplemental Material}
\twocolumngrid

\subsection*{Optical thickness measurement}
We employ three  different methods to measure the
optical thickness. First, we compute the theoretical transmission spectrum for various
$b_{\vbar}(\delta)$ values and use these profiles to fit the experimentally obtained transmission data. This leads to an optical
thickness  $b_{\vbar}(0) = 19$. Second, we
perform a shadow imaging experiment on the $^1S_0\,\rightarrow\,^1P_1$ broad transition ($\lambda_b
=\unit{461}{\nano\meter}$, linewidth $\Gamma_b = \unit{2\pi\times 32}{\mega\hertz}$), where Doppler broadening is
negligible. A collimated probe beam with a waist larger than the atomic cloud is sent onto the
cloud, and the transmission signal $\Iout/\Iin$ is measured using an electron multiplying CCD camera (\emph{Andor iXon
Ultra 897}). Typically, the probe frequency is set at a detuning, $\delta_b =\unit{53}{\mega\hertz}$, to reduce the
systematic error in the transmission measurement due to large optical thickness. The optical thickness $\mathcal{B}$ is
computed from the transmission signal, $\mathcal B = -\log (\Iout/\Iin)$, and is related to $b_0(0)$ of the intercombination
line by $b_0(0) = \mathcal B\left(1+4\delta_b^2/\Gamma_b^2\right) \lambda^2 / \lambda_b^2$. In our experiment,
we measure a peak value of $b_0(0)=95(5)$ using this method  and a corresponding value of
$b_{\vbar}(0) = 15(1)$ using $b_{\vbar}(0)=b_0(0)g(k\vbar/\Gamma)$.  Third, we carry out shadow imaging experiment directly on the
intercombination line transition. We vary the detuning in a range of 100 kHz around the resonance. The value of
$b_{\vbar}(0)$ is deduced using
\begin{equation}
   \Eout(\omega)=\Ein(\omega)\mathrm{e}^{i\frac{n(\omega)\omega L}{c}},
\end{equation}
and
\begin{equation}
   \alpha(\omega)=-\frac{3\pi\Gamma c^3}{\omega^3}\frac{1}{\sqrt{2\pi}\vbar}\int^{+\infty}_{-\infty}\ud v
      \frac{\mathrm{e}^{-v^2/2\vbar^2}}{\delta-kv+i\Gamma/2},
\end{equation}
which are~Eqs.~(1) and~(2) in the main text.
We have
$b_{\vbar}(0) = 19(2),$ a value slightly larger than the
one obtained by the second method.

\subsection*{Initial decay time $\tau_{\vbar}$}
We take $t=0$ as the time when the abrupt change occurs for the incident field
$\Ein$. To calculate the initial decay time of the cooperative forward transmitted field, we first note that we can rewrite~Eq.~(5) in the main text
as
\begin{equation}
   \tau_{\vbar} (\delta) = \left|\frac{1 - |\Eout(t=\infty)|^2
      / |\Eout(t=0^+)|^2}{2\re\left\{\frac{\D\Eout/\D t(t=0^+)}{\Eout(t=0^+)}\right\}}\right|,\label{eq:tau_field}
\end{equation}
where $\Eout(t=0^+) = \Ein(t=0^+)+\Esca(t=0^-)$, and $E_t(t=\infty)$ is the steady state transmitted field after the abrupt change in the incident field. For the case of abrupt extinction, $\Eout(t=0^+) = \Esca(t=0^-)$ and $\Eout(t=\infty) = 0$. For abrupt ignition, $\Eout(t=0^+) =\Ein$ and $|\Eout(t=\infty)| =|\Ein|\mathrm{e}^{-b/2}$. Here, $b=b_v(\delta)$ and $\theta=\theta_v(\delta)$, the same as defined in Eq.~(3) of the main text:
\begin{equation}
   b_{\vbar}(\delta) = \frac{2\omega}{c}\im[n(\omega)]L,\quad
   \theta_{\vbar}(\delta) = \frac{\omega}{c}\re[n(\omega)-1]L.
\end{equation}

 For abrupt phase jump by $\varphi$, we have $\Eout(t=0^+) =\Ein\mathrm{e}^{i\varphi}+\Esca(t=0^-),$ ignoring the small propagation time $L/c$ in the medium, and $|E_t(t=\infty)| =|\Ein| \mathrm{e}^{-b/2}$.
The forward scattered field during the steady state regime, in both cases of abrupt extinction and phase jump, is $\Esca(t=0^-) = \Ein\mathrm{e}^{-b/2+i\theta} - \Ein$.

In the denominator of~\eq{tau_field}, we need to compute the time derivative of the transmitted field at $t=0^+$. It can be computed by considering the derivative $\D\left[\Esca(t) \mathrm{e}^{i\omega t}\right]/\D t$. The forward scattered field in the time domain, $\Esca(t)$, is related to the incident field in the frequency domain, $\Ein(\omega')$, by the following well-behaved integral:


\begin{equation}
 \Esca(t)\\= \int \mathrm{e}^{-i\omega't} \left[\mathrm{e}^{i\frac{\omega'\rho\alpha\left(\omega'\right) L}{2c}}-1\right] \Ein(\omega')\D\omega'.
\label{eq:fs_field}
\end{equation}
The integration ranges of the integrals in this Supplemental Material, when not specified, are from $-\infty$ to $\infty$. $\Ein(\omega')$ is given for the cases of abrupt ignition, abrupt extinction and abrupt phase jump of $\varphi$ by:
\begin{equation}
\Ein(\omega') = \frac{i\xi \Ein}{2\pi}\mathrm{PV} \frac{1}{\omega'-\omega} + \frac{\eta \Ein}{2}\delta(\omega'-\omega).\label{eq:Eomega}
\end{equation}
where $\omega$ is the frequency of the probe. The Fourier variable corresponding to $t$ is denoted as $\omega'$.
 $\xi$ and $\eta$ are $-1$ and $1$ respectively for abrupt extinction of the probe, and $\mathrm{e}^{i\varphi}-1$ and $1+\mathrm{e}^{i\varphi}$ respectively for abrupt phase jump of the probe field. In the case of abrupt probe ignition, both $\xi$ and $\eta$ are equal to 1.
We substitute~\eq{Eomega} in~\eq{fs_field}, noting that the integral involving the Dirac delta function
goes to zero, to obtain
\begin{multline}
\frac{\D}{\D t} \left[\Esca(t) \mathrm{e}^{i\omega t}\right] \\= \frac{\xi \Ein}{2\pi}\sum_{p=1}^\infty \frac{1}{p!}\int \left(\frac{i\omega'\rho\alpha(\omega')L}{2c}\right)^p \mathrm{e}^{-i(\omega'-\omega)t}\D\omega'.
\end{multline}
We work in the regime where $\delta, \Gamma, k\vbar \ll \omega_0$. For $p=1$, the integral can be evaluated to be
\begin{multline}
\label{eq:toto}
\frac{\xi\Ein}{2\pi}\int \frac{i\omega'\rho\alpha(\omega')L}{2c} \mathrm{e}^{-i(\omega'-\omega)t}\D\omega'\\
 = \frac{\xi\Ein}{2\pi i} \frac{b_0(0)}{2}\frac{\Gamma}{2}\frac{1}{\sqrt{2\pi}\vbar}\iint \D v\,\D \omega' \frac{\mathrm{e}^{-i(\omega'-\omega)t} \mathrm{e}^{-v^2/2\vbar^2}}{\omega'-\omega_0-kv + i\Gamma/2}\\
 = -\xi\Ein\frac{b_0(0)\Gamma}{4}\mathrm{e}^{i\delta t}\mathrm{e}^{-\Gamma t / 2}\frac{1}{\sqrt{2\pi}\vbar}\int \D v \, \mathrm{e}^{-ikvt} \mathrm{e}^{-v^2/2\vbar^2}\\
 = -\xi\Ein\frac{b_0(0)\Gamma}{4}\mathrm{e}^{i\delta t}\mathrm{e}^{-\Gamma t / 2}\mathrm{e}^{-k^2\vbar^2 t^2 /2},
\end{multline}
for $t>0$. In~\eq{toto}, $b_0(0) = 6\pi \rho c^2 L/\omega_0^2$.  The $p>1$ terms, in general, are
difficult to evaluate for the general time dependence. Nevertheless, at $t=0^+$, they vanish. We
take the example of the term $p=2$, where essentially we have to deal with the following triple integral.
\begin{equation}
\iiint \frac{\mathrm{e}^{-i(\omega-\omega')t}\mathrm{e}^{-v^2/(2\vbar^2)}\mathrm{e}^{-v'^2/(2 \vbar^2)}\,\D \omega' \D v \,\D v' }{(\omega'-\omega_0-kv+i\Gamma/2)(\omega'-\omega_0-kv'+i\Gamma/2)}.
\end{equation}
We rewrite for $v\neq v'$,
\begin{multline}
\frac{\mathrm{e}^{-i(\omega-\omega')t}}{(\omega'-\omega_0-kv+i\Gamma/2)(\omega'-\omega_0-kv'+i\Gamma/2)}\\ = \frac{1}{k(v-v')}\frac{\mathrm{e}^{-i(\omega-\omega')t}}{\omega'-\omega_0-kv+i\Gamma/2}\\-\frac{1}{k(v-v')}\frac{\mathrm{e}^{-i(\omega-\omega')t}}{\omega'-\omega_0 - kv' + i\Gamma/2}.
\end{multline}
The integration over $\omega'$ of the above expression can be carried out easily, which results in 0 at $t=0^+$. When
$v=v'$, we have an integral over a multiple pole of order $2$, which also goes to 0 at $t=0^+$. Therefore, the term with
$p=2$ is zero at $t=0^+$. Similar argument can be extended to all orders $p>1$, showing that all $p>1$ terms vanish at $t=0^+$.
Finally, we have
\begin{equation}
\frac{\D}{\D t}\left[\Esca(t)\mathrm{e}^{i\omega t}\right] (t=0^+) = -\xi \Ein \frac{b_0(0) \Gamma}{4}.
\end{equation}
We then use the fact that $\Esca(t=0^+) = \Eout(t=0^+) - \Ein(t=0^+)$ to obtain
\begin{equation}
\frac{\D \Eout}{\D t} (t=0^+)  = - \xi \Ein \frac{b_0(0) \Gamma}{4} - i\omega \Eout(t=0^+).
\end{equation}
Using the above expression, we deduce the initial decay time for the case of abrupt probe extinction,
\begin{equation}
\tau_{\vbar}(\delta) = \frac{2}{\Gamma b_0(0)}\frac{1+\exp(-b)-2\exp(-b/2)\cos(\theta)}{1-\exp(-b/2)\cos(\theta)},
\end{equation}
which is Eq.~(5) of the main text. For the case of abrupt phase change, we find the initial decay time to be
\begin{equation}
\tau_{\vbar}(\delta) = \frac{4}{\Gamma b_0(0)} \frac{1-\exp(-b/2)\cos\theta}{2-\exp(-b/2)\cos\theta}.
\end{equation}
This is Eq.~(6) of the main text.
The initial decay time of the flash in the case of abrupt ignition is found to be
\begin{equation}
\tau_{\vbar}(\delta) = \frac{2}{\Gamma b_0(0)}\left[1-\mathrm{e}^{-b}\right].
\end{equation}
We observe again the appearance of the factor $2/\Gamma b_0(0)$ which arises from the cooperativity among the atomic dipoles.


In the case of an abrupt phase jump, we can further choose in the experiment, for $\varphi$ to be equal to the phase of $\Esca(t=0^-)$ relative to $\Ein(t=0^-)$. This choice ensures a constructive
interference after the phase jump. The decay time can be simplified to
\begin{equation}
   \tau_{\vbar}(\delta)= \frac{4}{\Gamma b_0(0)}\frac{|\Esca(t=0^+)|}{\Ein+|\Esca(t=0^+)|}.
\end{equation}

\end{document}